# Ictal and Post Ictal Impaired Consciousness due to Enhanced Mutual Information in Temporal Lobe Epilepsy

Puneet Dheer*, Systems Science and Informatics Unit, Indian Statistical Institute, 8th Mile, Mysore Road, Bangalore 560059, India; e-mail: puneetdheer@gmail.com

Sandipan Pati*, UAB Epilepsy Center, Department of Neurology, University of Alabama at Birmingham, CIRC 312, 1719 6th Avenue South, Birmingham, AL 35294, USA; e-mail: spati@uabmc.edu

Srinath Jayachandran, Systems Science and Informatics Unit, Indian Statistical Institute, 8th Mile, Mysore Road, Bangalore 560059, India; e-mail: sri9s@yahoo.in

Kaushik Kumar Majumdar, Systems Science and Informatics Unit, Indian Statistical Institute, 8th Mile, Mysore Road, Bangalore 560059, India; e-mail: kmajumdar@isibang.ac.in

*shared first authors

*Abstract*—Seizure and synchronization are related to each other in complex manner. Altered synchrony has been implicated in loss of consciousness during partial seizures. However, the mechanism of altered consciousness following termination of seizures has not been studied well. In this work we used bivariate mutual information as a measure of synchronization to understand the neural correlate of altered consciousness during and after termination of mesial temporal lobe onset seizures. First, we have compared discrete bivariate mutual information (MI) measure with amplitude correlation (AC), phase synchronization (PS), nonlinear correlation and coherence, and established MI as a robust measure of synchronization. Next, we have extended MI to more than two signals by principal component method. The extended MI was applied on intracranial electroencephalogram (iEEG) before, during and after 23 temporal lobe seizures recorded from 11 patients. The analyses were carried out in delta, theta, alpha, beta and gamma bands. In 77% of the complex partial seizures MI was higher towards the seizure offset than in the first half of the seizure in the seizure onset zone (SOZ) channels in beta and gamma bands, whereas MI remained higher in the beginning or in the middle of the seizure than towards the offset across the least involved channels in the same bands. Synchronization seems built up outside the SOZ, gradually spread and culminated in SOZ and remained high beyond offset leading to impaired consciousness in 82% of the complex partial temporal lobe seizures. Consciousness impairment was scored according to a method previously applied to assess the same in patients with temporal lobe epilepsy during seizure.

*Keywords*—**Consciousness score, beta-band, gamma-band, intracranial EEG (iEEG), mutual information, temporal lobe seizure.**

## 1. Introduction

Impaired consciousness in partial seizures is usually most profound late in the seizure and persists for up to several minutes after the termination of the seizure, during the post-ictal period (Blumenfeld and Taylor, 2003). Temporal lobe structures – hippocampus, amygdala and entorhinal cortex form heightened nonlinear correlation among themselves as well as with medial temporal gyrus, thalamus, posterior cingulate gyrus and lateral parietal cortex leading to moderate to profound alteration of consciousness during medial temporal lobe seizures (Arthuis et al., 2009). In that work a novel consciousness score was defined and was used to grade impairment of consciousness during temporal lobe seizures. This was then compared with a nonlinear correlation measure (Wendling et al., 2001) defined pairwise among the temporal lobe and extratemporal lobe structures mentioned above.

Neuronal gamma-band synchronization constitutes a fundamental process for all of cortical computations (Fries 2009). It is involved in modulating different aspects of consciousness (Ward, 2003), such as, visual awareness (Crick and Koch, 1990; Tallon-Baudry, 2009), face recognition (Rodriguez et al., 1999), associative learning (McIntosh et al., 1999), conscious perception (Melloni et al., 2007). Conscious thinking is often aided by working memory storage and long-term memory retrieval. Both the processes may interact in the hippocampus (Fell and Axmacher, 2011). An increase in coherence of theta and gamma oscillations in the hippocampus, amygdala and neocortex was predictive of immediate recall performance in a verbal learning task (Babiloni et al., 2009).

The brain network that is involved in a cognitive process has to be reactivated when the cognitive task is to be replayed. The cognitive process is modulated by beta-band synchronization across the network (Spitzer and Haegens, 2017), and is important for endogenous content retrieval during conscious thoughts. Alpha-band synchronization has been implicated in suppressing distracting stimuli in order to pay enhanced attention to the desired one (Ward, 2003). It has also been implicated in working memory. Gamma-band synchronization for feature binding may require alpha-band desynchronization (Ward, 2003). Experimental data concerning event-related theta oscillations hint at a basic role in cognitive processing and in the cortico-hippocampal interaction. Therefore excessive theta-band synchronization in the temporal lobe may cause impaired cognition. Stronger theta and delta synchronization has been implicated in the processing of stimuli with higher emotional value over emotionally neutral stimuli (Knyazev et al., 2009). Delta and theta responses have also been implicated in arousal (Guntekin and Basar, 2016). It is suggested that complex stimuli elicit superimposed alpha, gamma, theta responses which are combined like letters in an alphabet (Basar et al., 2001). In general, conscious awareness may arise from synchronous neural oscillations occurring globally throughout the brain rather than from the locally synchronous oscillations that occur when a sensory area encodes a stimulus (Ward, 2003). It is natural that during epileptic seizures excessive synchronization in the form of a nonlinear correlation (Wendling et al., 2001) across all the frequency bands overloaded the brain structures involved in conscious processing leading to impaired consciousness (Arthuis et al., 2009).

Synchronization is a ubiquitous phenomenon in neuroscience. It is present from cellular level (Steinmetz et al., 2000) to systems level in primate brains (Ward, 2003). However, even within neuroscience synchronization is a generic term, interpreted and measured differently in different contexts. Sometimes it is phase synchronization (Rodriguez et al., 1999), some other time it is plain linear amplitude correlation (Schindler et al., 2007) and yet other times it is nonlinear correlation (Arthuis et al., 2009). Mean phase coherence (Mormann et al., 2000) or phase locking value (Aydore et al., 2013) is another widely used measure of synchronization in brain networks. Whatever be the measure, synchronization looks for simultaneity in processing by different parts of the brain network for a cognitive task. In general, this simultaneity may be with any time lag (Arthuis et al., 2009). If different parts of the brain are considered different dynamical systems, each of which is manifested through a time series generated by the underlying dynamical system, then measuring synchronization across the brain network boils down to finding simultaneous events or patterns in the time series.

In neuroscience, be it electrophysiological signals, magnetic signals, hemodynamic activation time series, ionic concentration variation time series, optical time series, or infrared time series, it is the geometric shape which uniquely determines the time series like a face or fingerprint. All the information embedded in the signal are embedded in its shape. Sometime this information is extracted as phase, sometime it is extracted as amplitude, sometime it is extracted as frequency, etc. One or more of the features together represent partial or total information encoded in the time series. Recently it has been shown that information that gives meaning to the time domain signal (not the statistical properties of information as in Shannon theory) and analyzed during processing of the physiological signal for diagnostic and research purposes is encoded in 3-point neighborhoods of the digital signal (Majumdar and Jayachandran, 2018). Such information is encoded in ordinal patterns generated by permutations of the 3 points in the neighborhood (Olofsen et al., 2008). These ordinal patterns have been utilized to measure permutation conditional mutual information (PCMI) between two neural signals (X. Li and G. Ouyang, 2010).

Study of mutual information across multiple intracranial EEG (iEEG) channels in temporal lobe across delta, theta, alpha, beta and gamma bands before, during and after epileptic seizures vis-à-vis the state of consciousness of the patient may offer us important insights into the relationship between seizure and consciousness. Impaired consciousness during seizure and in the post-ictal state contributes to significant mortality and morbidity in patients with epilepsy. Moreover, if mutual information can be shown to be a more general form of synchronization, correlation or coherence measure than phase synchronization or amplitude correlation or other simultaneity measures mentioned above, the conclusions drawn with the help of mutual information measure will be more general in nature than achievable by any one single simultaneity measure.

## 2. Methods

Measures like correlation, coherence, synchronization, etc are used to estimate the interdependence between two signals. Mutual information calculates this interdependence in terms of an information distance measure, called Kullback-Leibler divergence (Cover and Thomas, 2006) from the joint distribution of the two random variables (signals) to the product of the marginal distributions. Mutual information $I(x, y)$ between two time domain signals $x$ and $y$ is given by

$$I(x, y) = \sum_x \sum_y p(x, y) \log \frac{p(x, y)}{p(x) p(y)}, \tag{1}$$

for discrete time signals $x$ and $y$. Clearly, when $x$, $y$ are independent $I(x, y) = 0$, or, $I(x, y) > 0$ otherwise.

It is important to note that correlation implies mutual information. For example, mutual information between two correlated Gaussian random variables with correlation coefficient 1 is infinite, and with correlation coefficient 0 the mutual information between them becomes 0 (Cover and Thomas, 2006, p. 252). In Fig. 1 we have presented a comparison among various synchronization measures prevalent in Neuroscience. Since synchronization has been studied most extensively during epileptic seizures, we have presented our results using different synchronization measuring algorithms on a pair of human iEEG signals (unfiltered) collected before, during and after an epileptic seizure.

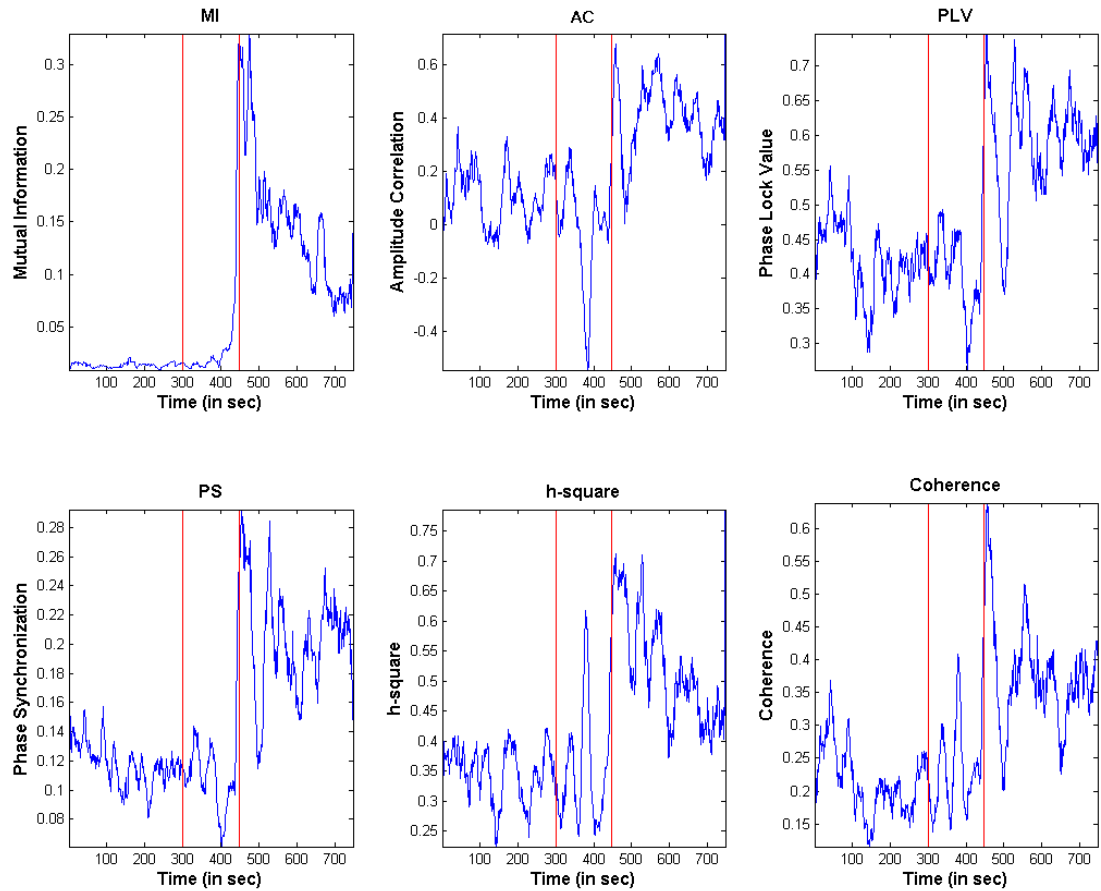

l

Fig. 1. Comparison among Mutual Information (MI), Amplitude Correlation (AC), Phase Locking Value (PLV) (Aydore et al., 2013), (Hilbert transformation based) Phase Synchronization (PS), Nonlinear Correlation or h-square value (Wendling et al., 2001) and coherence (Bastos and Schoffelen 2016) over a pair of pre-ictal, ictal and post-ictal signals. Vertical lines indicate seizure onset and offset time points.

Synchronization during seizure is a complex, heterogeneous process over the time and space (Jiruska et al., 2013), which has been captured by the methods in Fig. 1. However, synchronization is the highest towards the end of seizure which may lead to spontaneous termination (Majumdar et al., 2014), and that also has been clearly captured. Mutual information may be high due to one or more of the simultaneity measures, only some of which have been shown in Fig. 1. Some of the measures in Fig. 1 show conflicting trends till before the end of the seizure and that could be the reason why MI value is smaller all along than any of the other measures and capturing the high synchronization trend during the offset in a sharper way than the other methods, when they all concur.

Apart from equation (1), MI can be interpreted in another way. Entropy is the uncertainty of a single random variable. We can define conditional entropy $H(x \mid y)$, which is the entropy of a random variable conditional on the knowledge of another random variable. The reduction in uncertainty due to another random variable is called the mutual information. For two random variables $x$ and $y$ this reduction is the mutual information

$$I(x,y) = H(x) - H(x \mid y) = \sum_x \sum_y p(x,y) \log \frac{p(x,y)}{p(x)p(y)} \,. \tag{2}$$

The mutual information $I(x,y)$ is a measure of the dependence between the two random variables. It is symmetric in $x$ and $y$ and always nonnegative and is equal to zero if and only if $x$ and $y$ are independent (Cover and Thomas, 2006, p. 6-7).

Likewise we can say – amplitude correlation is a condition under which reconstruction of amplitude of one signal needs conditional knowledge of amplitude of the other signal, provided that there are only two signals. In mathematical formulation it will appear as

$$I(A(x), A(y)) = H(A(x)) - H(A(x) \mid A(y))\,, \tag{3}$$

where $A(\ )$ denotes the amplitude. Similarly, for phase synchronization we can write

$$I(Ph(x), Ph(y)) = H(Ph(x)) - H(Ph(x) \mid Ph(y))\,, \tag{4}$$

where $Ph(\ )$ denotes the phase. Likewise, this can be extended to any measure of simultaneity or interdependence between two signals. MI between $x$ and $y$ is then given by

$$I(x,y) = \bigcup_S I(S(x), S(y))\,, \tag{5}$$

where $S$ is a suitable feature of the signal for measuring simultaneity or interdependence with another signal.

In case of a pair of discrete signals MI can be measured in different ways. Here we have measured MI in terms of motifs in time series (Lin et al., 2002). Information is encoded in a time series in terms of three point motifs or 3-motifs (Majumdar and Jayachandran, 2018). 3-motifs, generated by permutation of the 3 values, have been utilized to extract information out of EEG signals (Olofsen et al., 2008). Those $3! = 6$ motifs have been utilized to measure Permutation Conditional Mutual Information (PCMI) between two signals (Li and Ouyang, 2010). PCMI gives a directional measure of how one signal is dependent on the other.

Here we have utilized the same 3-motifs to measure MI between two discrete signals (Fig. 2). Since there are 6 number of 3-motifs altogether let us number them from 1 to 6. Take a window of length, say 1000 samples in both the signals $x$ and $y$, starting and ending at the same times across the signals. There are a total of 998 3-motifs in the window from each of the signals. Say, the first motif in $x$ is the motif number 1 and the first motif in $y$ is the motif number 5 (Fig. 2). Then in the $2 \times 998$ signal motif matrix the first column will be $\begin{bmatrix} 1 \\ 5 \end{bmatrix}$. The frequency density of $\begin{bmatrix} 1 \\ 5 \end{bmatrix}$ over the 998 motif window will be $p(1,5)$. Similarly, if we consider the frequency density of all the columns of the signal motif matrix, we will eventually get $p(x,y)$ and will be able to calculate $I(x,y)$ from (1). $p(x)$, $p(y)$ and $p(x,y)$ have been shown in Fig. 3.

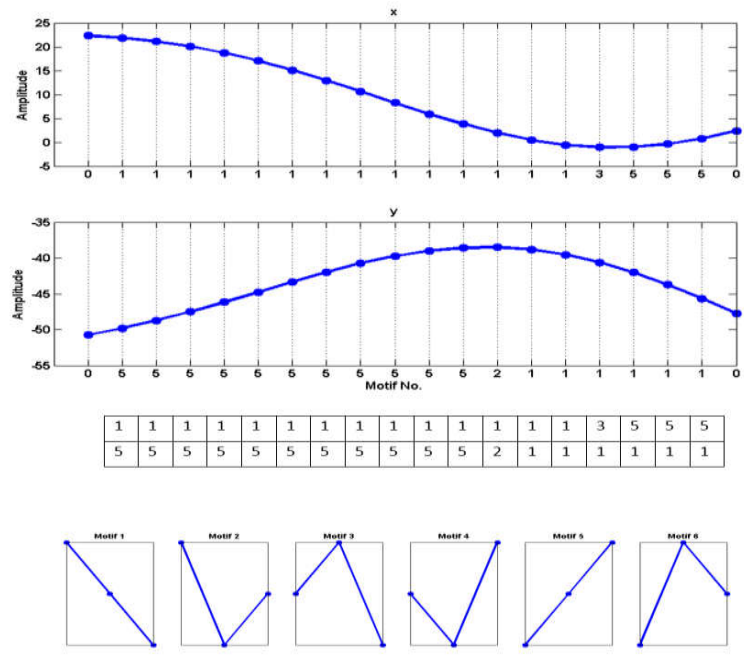

Fig. 2. Motif decomposition of discrete signals $x$ and $y$ (first two plots in the top). Signal motif matrix in the thirds row from the top, and all 6 3-motifs numbered at the bottom. These are the same motifs as in (Olofsen et al., 2008) and (Li and Ouyang, 2010).

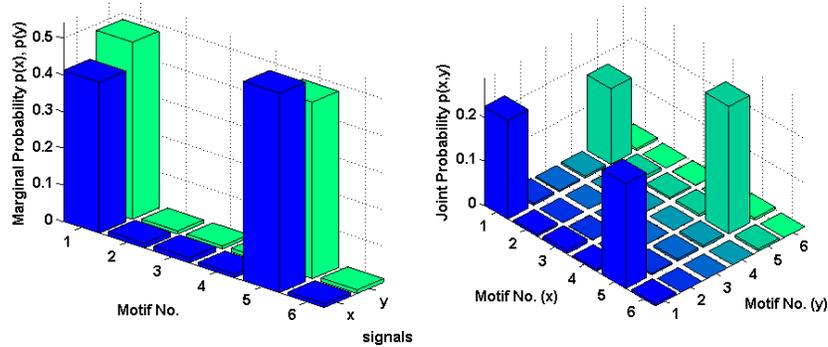

Fig. 3. On the left, $p(x)$ as frequency density of 6 3-motifs in $x$ has been shown in blue histogram and the same for $y$ as $p(y)$ has been shown in green histogram. On the right, the joint probability density function $p(x, y)$ has been plotted.

For extension of this measure from two signals to more than two signals we have used the procedure of (Schindler et al., 2007). This extension was done to generalize the correlation measure between two epileptic iEEG signals to more than two epileptic iEEG signals during seizures. Since here our purpose is the same, we have chosen this extension procedure to generalize MI between two signals to more than two signals. If there are $n$ number of iEEG channels we form the $n \times n$ matrix, whose $ij$ th entry is $I(x_i, x_j)$ over a window, where $x_i$ is the iEEG signal collected from the $i$ th channel. This matrix is a symmetric matrix and therefore all eigen values are real. By sliding the window we take only the highest eigen value from each of the windows. The plot of the highest eigen value over the time windows will give the dominant trend of MI across all the $n$ channels. This is akin to the principal component analysis. For a high value of MI across all $n$ channels, not all channels will necessarily have a strong overlap. Even if a few of them have strong overlap and others do not have as strong overlaps, the MI across all $n$ of them can still be high. Unfortunately, this algorithm cannot identify which subset of channels has strong overlap, say, above a particular threshold.

### 3. Data

23 seizures from 11 patients have been studied. Detail of the patients and the seizures have been furnished in Table 1. Consciousness was impaired if the consciousness score is greater than 2 (Table 2). Selection of patients and intracranial EEG recordings: Of the 42 adults with drug-refractory focal epilepsy who have undergone intracranial EEG investigation to localize seizures at the University of Alabama at Birmingham, AL level - IV epilepsy center between 2014 and 2017, we have selected 11 patients in this study. The patient selection criteria were as following: a) seizure onset was within the mesial temporal lobe structures (amygdala, hippocampus, parahippocampal gyrus, temporal pole); b) have undergone anterior temporal lobectomy; c) had at least 6 months post resection follow up; and d) seizure outcome was favorable (Engel I and II) (Engel et al. 1993). See Table 3 for post operative prognosis of all the 11 patients. All except one (patient 7) had previous failed epilepsy surgery due to incomplete resection of mesial temporal lobe structures. All patients had undergone standard investigation before intracranial EEG study, and this includes 3T MRI brain, 18-fluorodeoxy glucose PET scan, Magnetoencephalography and scalp EEG investigation. Of the eleven subjects, ten had robot-assisted stereo EEG investigation with sampling from the anterior and posterior hippocampus, amygdala, temporal pole, superior temporal gyrus, orbitofrontal, anterior insula, anterior cingulate and parahippocampal gyrus. Each depth electrode had 10 – 14 contacts (PMT electrodes, 1.4 mm contacts). In one subject (subject 4) grids, strips and a depth electrode were placed for electrocorticography targeting hippocampus, subtemporal and lateral temporal regions. Scalp EEG was recorded from Cz, Pz. Video EEG was sampled at around 2 KHz using Natus Xltek EEG machine (see Table 1 for detail of sample frequency). Post explantation all video EEG and other investigations were discussed in a multidisciplinary epilepsy surgical conference to confirm the onset of seizures and seizure onset channels. These are standard procedures followed in any level – IV epilepsy centers. To compare mutual information findings with the seizure onset zone channels with channels that had late seizure propagation, we have classified these channels as late (or least) involved channels (LIC). We have selected these channels by visual interpretation of EEG in a systematic way as described below.

Initially, we have identified seizure termination and then visually analyzed EEG from all recorded channels moving backward towards seizure onset. For focal seizures without generalization, we have selected LIC channels that had no ictal epileptiform activity on visual interpretation. For focal seizures that were generalized, we selected LIC channels that had late (or last) propagation of ictal epileptiform activity. The post operative outcome was obtained from retrospective chart review. The study was approved by the institutional review board.

Selection of seizures and scoring of consciousness: 23 seizures in 11 patients have been studied. Detail of the patients and the seizures have been furnished in Table 1. Selection criteria for seizures were as follows- a) only spontaneous seizures were selected; b) video of the seizure was available; c) electrographic seizures that lacked any clinical accompaniment were excluded; d) nursing staff was able to examine the patient during and after the seizure. In our epilepsy center we have adopted a protocol to examine the patient during and after seizure by the nursing staffs. The examination includes : a) evaluating awareness by interaction with patient ( asking name, place, date); b) visual attention by seeing if they follow them or an object (such as, pen), during conversation; c) motor examination by asking them to lift arms, legs d) speech/language by asking them to read from a board and e) asking patient to report what they feel during seizure. These examinations were continued in the post ictal period and at times continued periodically till the patient was back to baseline. By reviewewing the video , one of the authors scored consciousness following termination of seizure. This scale takes into account different aspects of consciousness in humans: (i) unresponsiveness (Criteria 1 and 2); (ii) visual attention (Criteria 3); (iii) consciousness of the seizure (Criteria 4); (iv) adapted behavior (Criteria 5); and (v) amnesia (Criteria 6 and 7) (Arthuis et al. 2009).

Table 1: Summary of patient demographics and seizures

| Sz No. | Age | Gender | Onset | Offset | SOZ Channels | Pre sz vigilance | Sz | Post sz behavioral change |
|---|---|---|---|---|---|---|---|---|
| **1** | 23 | F | 300 | 580 | Left Hpc, PHG | awake | SP | Awake, alert |
| **2** | | | 300 | 482 | Left Hpc, PHG | awake | CP | Awake, confused |
| **3** | 19 | M | 300 | 480 | Right Amyg Hpc | awake | CP | Awake, confused |
| **4** | | | 300 | 473 | Right Amyg Hpc | awake | CP | Awake, confused |
| **5** | | | 300 | 377 | Right Amyg Hpc | awake | CP | Awake, confused |
| **6** | 38 | M | 300 | 371 | Right Hpc, PHG | sleep | sGTC | Awake, confused |
| **7** | | | 300 | 482 | Right Hpc, PHG | awake | CP | Awake,  confused |
| **8** | 21 | F | 300 | 467 | Left PHG, STG, MTG, | Stg II sleep | sGTC | Awake, confused |
| **9** | | | 300 | 375 | Left PHG, STG, MTG | Stg II sleep | CP | Awake, confused |
| **10** | | | 300 | 351 | Left PHG, STG, basal temp | awake | CP | Awake, confused |
| **11** | 25 | F | 300 | 389 | Left Amyg, Hpc, Tp | awake | CM | Awake, confused |
| **12** | | | 300 | 419 | Left Amyg, Hpc, Tp | awake | sGTC | Obtunded, minimal response |
| **13** | 42 | M | 300 | 352 | Right Hpc, PHG, basal temp | awake | sGTC | Awake, confused |
| **14** | | | 300 | 390 | Right Hpc, PHG, basal temp | awake | sGTC | Awake, confused |
| **15** | 24 | F | 300 | 434 | Left mesial temp remnant | awake | CP | Awake, confused |
| **16** | | | 300 | 400 | Left mesial temp remnant | awake | CP | Awake, confused |
| **17** | 44 | M | 300 | 383 | Right Hpc, PHG, basal temp | Stg II | sGTC | Obtunded, |

| | | | | | | sleep | | minimal response |
|---|---|---|---|---|---|---|---|---|
| **18** | | | 300 | 388 | Right PHG, basal temp , | Stg II sleep | sGTC | Obtunded, minimal response |
| **19** | 37 | M | 300 | 367 | Left Hpc, PHG, basal temp | awake | CP | Awake, confused |
| **20** | | | 300 | 363 | Left Hpc, PHG, basal temp | awake | CP | Awake, confused |
| **21** | 22 | M | 300 | 455 | Left Hpc, Amyg | awake | CP | Awake, confused |
| **22** | | | 300 | 431 | Left Hpc, Amyg | awake | CP | Awake, confused |
| **23** | 19 | M | 300 | 447 | Right Hpc, Amyg | awake | CP | Awake, confused |

Note: Sample frequency is 2048 Hz except for seizures 11, 12 (2000 Hz each) and 19, 20 (1000 Hz each). Hpc = Hippocampus, PHG = Parahippocampal gyrus, Amyg = Amygdala, STG = Superior temporal gyrus, MTG = Medial temporal gyrus, Tp = Temporal pole, SP = Simple partial seizure (new classification focal seizure with awareness), CP = complex partial (new classification focal seizure with impaired awareness), sGTC = secondary generalized tonic-clonic (new classification focal to bilateral tonic-clonic).

## 4. Results

a. Progressive increase in MI toward seizure termination and in post ictal period: As complex partial seizures progress from onset to termination, there is an increase in MI across all frequency bands (delta, theta, alpha, beta and gamma) and across all seizure onset zone (SOZ) channels (the most involved channels) and all least involved channels (LIC, Fig. 4). An example of this is represented in Fig. 5 (seizure 20 of patient 9), where consciousness score (Arthuis et al. 2009) after the seizure offset is 6. This means the consciousness is significantly impaired in the post ictal periods and this is expressed by the state awake/confused (Table 1). Although the MI is increased towards offset in all frequency bands (delta to gamma), the trend is weakest in delta and strongest in gamma band. Furthermore, high MI towards offset trend is sharper in SOZ channels than the LIC. Both the trends remained preserved through all 23 seizures (Supplement). These important trends were observed with MI were more distinct than the trends plotted with AC, PLV, PS, h-square (Fig. 1). If the seizure duration is subdivided into two equal parts then MI across the SOZ channels is higher in the second half than in the first with the highest degree of statistical significance among MI, AC, PLV, PS and h-square (Table 4, Table 5, Table 6, Table 7), except PS has marginally better significance in gamma band.

Table 2: Seizure duration and total consciousness score after seizure termination

| Sz No. | Sz duration in sec (#SOZ channels, #LIC) | Total Score |
|---|---|---|
| 1 | 280 (12, 31) | 0 |
| 2 | 182 (12, 31) | 3 |
| 3 | 180 (6, 12) | 3 |
| 4 | 173 (6, 12) | 1 |
| 5 | 77 (6, 12) | 2 |
| 6 | 71 (7, 6) | 4 |
| 7 | 182 (7, 6) | 4 |
| 8 | 167 (10, 12) | 6 |
| 9 | 75 (10, 12) | 2 |
| 10 | 51 (10, 12) | 8 |
| 11 | 89 (12, 16) | 7 |
| 12 | 119 (12, 16) | 9 |
| 13 | 52 (10, 5) | 4 |
| 14 | 90 (10, 5) | 9 |
| 15 | 134 (12, 14) | 1 |
| 16 | 100 (12, 14) | 5 |
| 17 | 83 (6, 6) | 8 |
| 18 | 88 (6, 6) | 7 |
| 19 | 67 (5, 7) | 4 |
| 20 | 63 (5, 7) | 6 |
| 21 | 155 (11, 26) | 6 |
| 22 | 131 (11, 26) | 4 |
| 23 | 147 (8, 6) | 4 |

Note: Total number of implanted channels per patient varies from 150 to 240. SOZ = seizure onset zone, LIC = least involved channels.

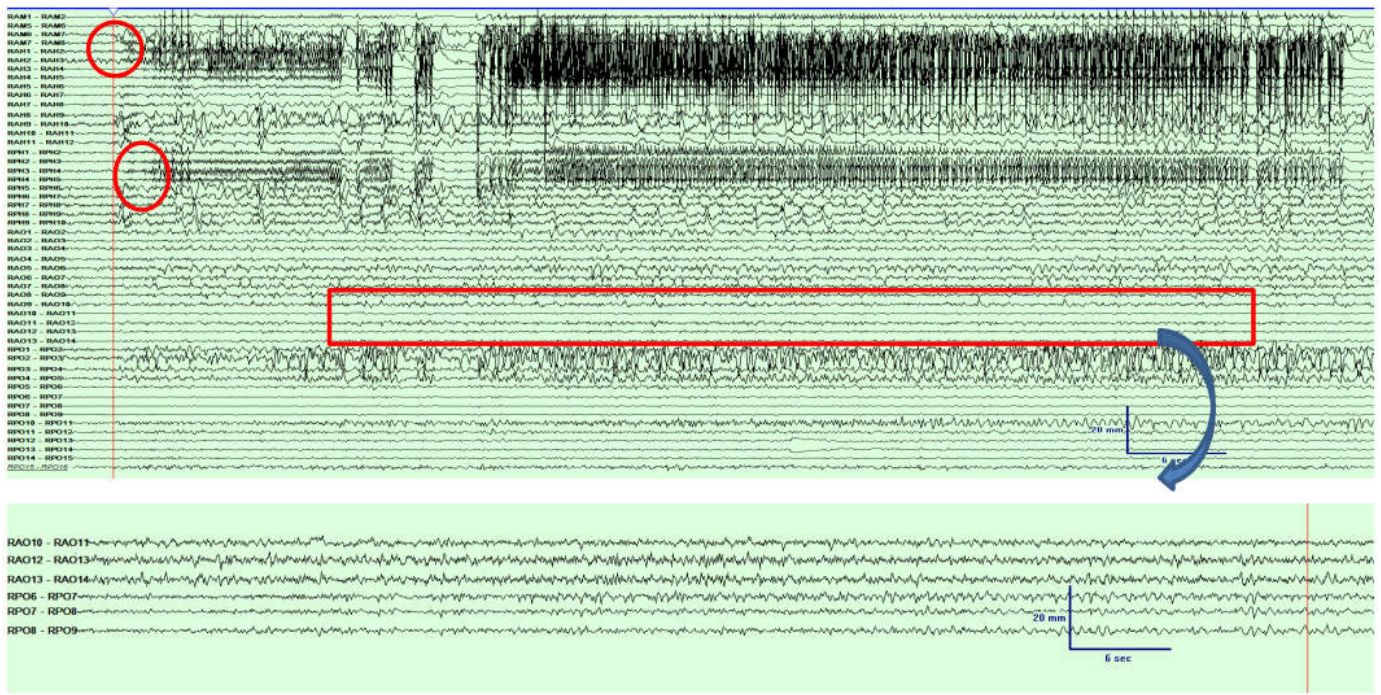

Fig. 4. Electrical activities in seizure onset zone channels and least involved channels. Red circles indicate seizure onset patterns in iEEG recording. Red vertical line in the top plot indicates seizure onset time. Red rectangle indicates the activities in the least involved channels, which are magnified in the bottom plot.

b. Comparison of MI with other measures of synchronization: Each of the 11 patients was implanted with 150 to 240 iEEG electrodes for evaluation of possible surgical resection. High degree of synchronization (measured by h-square) across temporal lobe has been implicated in loss of consciousness during temporal lobe seizures (Arthuis et al., 2009). Since multidimensional synchronization is difficult to measure only a few pairwise synchronization has been studied in (Arthuis et al., 2009). We have extended pairwise analysis of MI to higher dimension following a previous published study (Schindler et al., 2007). This analysis is not applicable to h-square because it is not a symmetric measure. However, if we are to measure MI across all the 150 to 240 channels we will have to calculate $n \times n$ matrix multiplication for $150 \le n \le 240$, for each time window across the $n$ iEEG signals. The computation becomes so huge that for little more than 100 channels that dual Xeon E5-2620v3 processor based GPU workstation (Nvidia K40 Tesla card) with 256 GB RAM ran out of memory midway through the computation (executed through MATLAB). So, we kept our analysis confined within the most involved channels (the seizure onset zone channels) and least involved channels (Fig. 4) in each seizure (this was determined by the author SP, who is an epileptologist). We also arbitrarily added other channels in the above two sets, but the trend didn't change (Fig. 5). From this we conclude that MI remained high across the channels towards the offset of the seizures particularly in the higher frequency bands implicated in higher cognitive functions and consciousness.

For normal functioning of the brain synchronization will have to be at a normal level. Too high or too low a level of synchronization is pathological. We found support for this hypothesis through the multidimensional MI measure (Table 4). MI is higher in beta and gamma bands in the second half of the seizures (towards the offset) across the SOZ channels ( $p < 0.05$ ). The same finding in all other frequency bands is not statistically significant as can be seen in Table 4. Multidimensional extension of MI was done flowing (Schindler et al. 2007) by highest eigenvalue of the MI square matrix. The same extension technique was followed for all other symmetric synchronization measure presented in Table 5, Table 6 and Table 8. In all the Tables from 4 through 8 the p value is for the hypothesis that the synchronization value will be higher during the second half of the seizure than in the first.

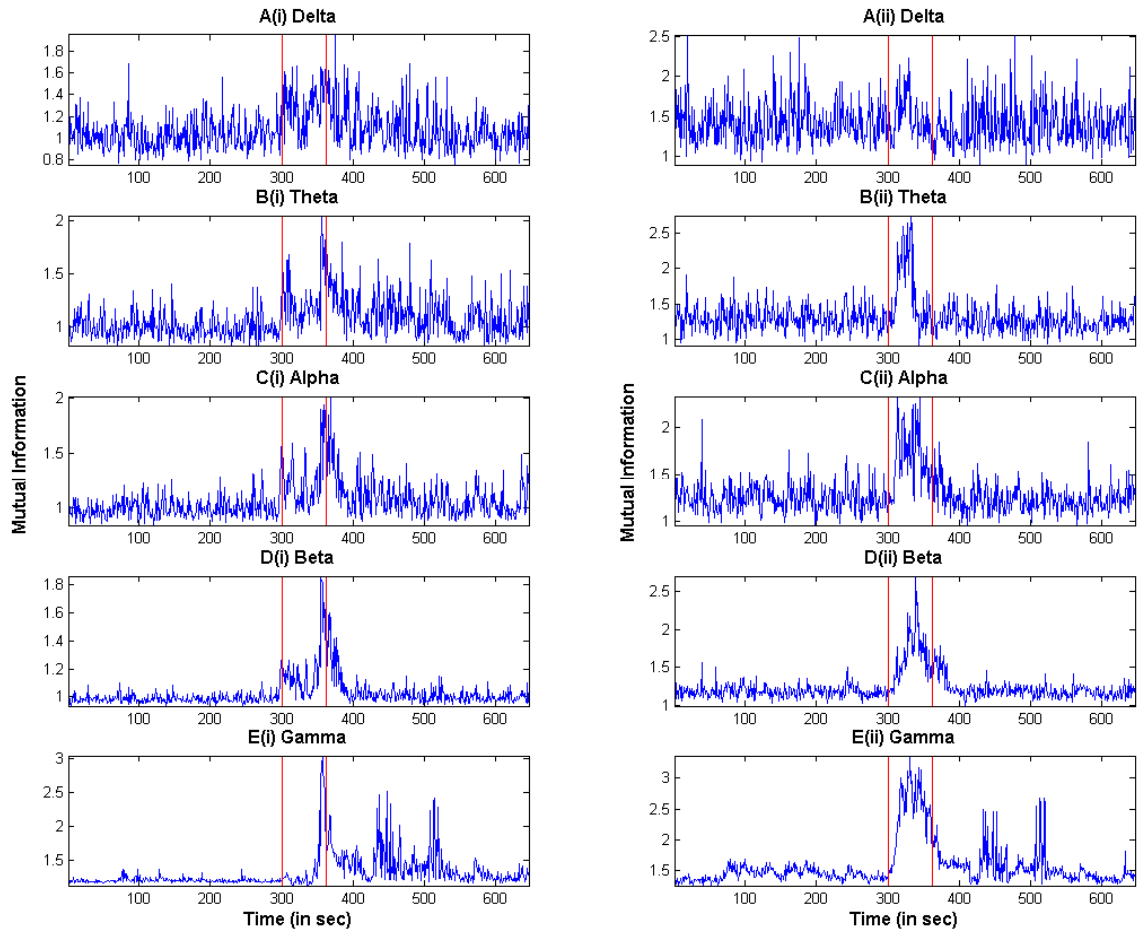

Fig. 5. Mutual information across all the 5 electrodes from the SOZ (left column) and all the 7 LIC (right column) of seizure 20 in Table 1, in the frequency bands of delta $(0-4$ Hz$)$ (A(i) for SOZ and A(ii) for LIC), theta $(4-8,$ Hz$)$ (B(i) for SOZ and B(ii) for LIC), alpha $(8-12$ Hz$)$ (C(i) for SOZ and C(ii) for LIC), beta $(12-30$ Hz$)$ (D(i) for SOZ and D(ii) for LIC) and gamma $(30-80$ Hz$)$ (E(i) for SOZ and E(ii) for LIC). Red vertical lines indicate seizure onset and offset time.

Table 3: Post operative prognosis of the patients

| Subjects | Epilepsy duration (years) | Aura | LOC (Y/N) | Surgery | Follow-up (months) | Outcome | Histopathology |
|----------|---------------------------|------|-----------|---------|--------------------|---------|----------------|
| 1 | 6 | Experiential feeling, Deja vu | Y | Left ATL | 9 | I | HS |
| 2 | 4 | Warm feeling over chest | Y | Right ATL | 8 | II | Granulomatous inflammatory |

| 3 | 35 | Déjà vu, epigastric | Y | Right ATL | 14 | II | Gliosis, FCD I |
| 4 | 9 | Abnormal sensation in head, experiential feeling | Y | Left ATL extended to lateral temporal | 12 | I | Gliosis |
| 5 | 2 | Anxious, intense fear | Y | Left ATL | 16 | I | Gliosis |
| 6 | 31 | Deja vu | Y | Right ATL | 14 | II | HS |
| 7 | 23 | Dizziness, experiential feeling | Y | Left mesial residual structures | 7 | II | Gliosis |
| 8 | 38 | Déjà vu. | Y | Right ATL | 18 | I | HS |
| 9 | 12 | Déjà vu, anxious feeling | Y | Left ATL | 20 | I | HS |
| 10 | 3 | Experiential feeling | Y | Left ATL | 14 | II | Non specific |
| 11 | 2 | Deja vu | Y | Right ATL | 6 | I | HS |

ATL= Anterior temporal lobectomy; FCD = Focal cortical dysplasia; HS = Hippocampal sclerosis; LOC = Loss of consciousness; Outcome I and II pertain to Engel class I and Engel class II epilepsy respectively.

Table 4: Statistical significance values of MI being higher during the second half of the seizure than the first according to Wilcoxon Rank-Sum test across all 23 seizures of 11 patients

| Frequency band | SOZ channels | LIC |
|---|---|---|
| Delta (0 − 4 Hz) | p=0.7584 (>0.05) | p=0.9125 (>0.05) |
| Theta (4 − 8 Hz) | p=0.7752 (>0.05) | p=0.6604 (>0.05) |
| Alpha (8 − 12 Hz) | p=0.6604 (>0.05) | p=0.9125 (>0.05) |
| Beta (12 − 30 Hz) | p=0.0331 (<0.05) | p=0.6764 (>0.05) |
| Gamma (30 − 80 Hz) | p=0.0026 (<0.05) | p=0.3795 (>0.05) |

Interestingly, gamma (Fries, 2009) and beta (Spitzer and Haegens, 2017) rhythms have been implicated mostly in conscious information processing by the brain. It has been shown that fast activity (at the beta range) across both hemispheres in the temporal lobe along with slow (1 − 2 Hz) activity in the fronto-parietal region leads to loss of consciousness during complex partial temporal lobe seizures (Englot et al., 2010). Clearly, beta band synchronization across the temporal lobe is likely to be high during the seizure. We have specifically shown that MI (in Table 4) and h-square (Wendling et al. 2001) (in Table 7) are high across all the SOZ channels during the second half of the temporal lobe seizures in beta band with $p < 0.033$ and $p < 0.002$ respectively (since h-square is not a symmetric measure, its extension from two channels to more than two channels is different from the other bivariate measures presented in Table 4, Table 5, Table 6 and Table 8). The same is true for gamma band with $p < 0.026$ and $p < 0.002$ respectively. Gamma band higher synchronization is statistically significant ($p < 0.010$) also by PS measure across the SOZ channels in the second half of the seizure than in the first (Table 5).

Table 5: Statistical significance values of PS being higher during the second half of the seizure than the first according to Wilcoxon Rank-Sum test across all 23 seizures of 11 patients

| Frequency band | SOZ channels | LIC |
|---|---|---|
| Delta (0 – 4 Hz) | p=0.3562 (>0.05) | p=0.7088 (>0.05) |
| Theta (4 – 8 Hz) | p=0.5385 (>0.05) | p=0.5828 (>0.05) |
| Alpha (8 – 12 Hz) | p=0.9475 (>0.05) | p=0.6134 (>0.05) |
| Beta (12 – 30 Hz) | p=0.1471 (>0.05) | p=0.8433 (>0.05) |
| Gamma (30 – 80 Hz) | p=0.0108 (<0.05) | p=0.5385 (>0.05) |

Table 4 through Table 8 show that among the bivariate measures of synchrony like MI, PS, AC, h-square (Wendling et al. 2001) and coherence (Bastos and Schoffelen 2016), extended to most involved (SOZ channels) and least involved channels, only MI and h-square show statistically significant synchronization across the most involved channels in beta and gamma range. Statistically significant synchronization across all the 23 seizures of 11 patients in beta and gamma band supports the finding of (Arthuis et al. 2009) that, loss of consciousness happens during temporal lobe seizures due to wider synchrony across brain regions involved in processing of awareness. Gamma synchronization takes place at a smaller spatial scale and beta synchronization takes place at a larger spatial scale (Kopell et al. 2000). When both of them occur with high significance across the seizure onset zone towards seizure termination and in post ictal state, that indicates local gamma synchronized clusters (the small localized brain regions oscillating at gamma rhythm) have become synchronized among themselves across wider brain regions leading to impairment of consciousness (Arthuis et al. 2009).

Table 6: Statistical significance values of AC being higher during the second half of the seizure than the first according to Wilcoxon Rank-Sum test across all 23 seizures of 11 patients

| Frequency band | SOZ channels | LIC |
|---|---|---|
| Delta (0 − 4 Hz) | p=0.8951 (>0.05) | p=0.9475 (>0.05) |
| Theta (4 − 8 Hz) | p=0.8261 (>0.05) | p=0.6134 (>0.05) |
| Alpha (8 − 12 Hz) | p=0.6604 (>0.05) | p=0.8090 (>0.05) |
| Beta (12 − 30 Hz) | p=0.3795 (>0.05) | p=0.9300 (>0.05) |
| Gamma (30 − 80 Hz) | p=0.2916 (>0.05) | p=0.5385 (>0.05) |

Table 7: Statistical significance values of h-square (Wendling et al. 2001) being higher during the second half of the seizure than the first according to Wilcoxon Rank-Sum test for all 23 seizures of 11 patients

| Frequency band | SOZ channels | LIC |
|---|---|---|
| Delta (0 − 4 Hz) | p= 0.2916 (>0.05) | p=0.9475 (>0.05) |
| Theta (4 − 8 Hz) | p= 0.2423 (>0.05) | p=0.3229 (>0.05) |
| Alpha (8 − 12 Hz) | p= 0.0054 (<0.05) | p=0.2720 (>0.05) |
| Beta (12 − 30 Hz) | p= 0.0024 (<0.05) | p=0.3916 (>0.05) |
| Gamma (30 − 80 Hz) | p= 0.0023 (<0.05) | p = 0.9825 (>0.05) |

Note: Since h-square is a non-symmetric measure between two channels, its multidimensional extension was done by taking average across all the pairwise values.

Table 8: Statistical significance of coherence values (Bastos and Schoffelen 2016) being higher during the second half of the seizure than the first according to Wilcoxon Rank-Sum test across all 23 seizures of 11 patients

| Frequency band | SOZ channels | LIC |
|---|---|---|
| Delta (0 − 4 Hz) | p=0.5531 (>0.05) | p=0.9650 (>0.05) |
| Theta (4 − 8 Hz) | p=0.9650 (>0.05) | p=0.6134 (>0.05) |
| Alpha (8 − 12 Hz) | p=0.3916 (>0.05) | p=0.7752 (>0.05) |
| Beta (12 − 30 Hz) | p=0.2269 (>0.05) | p=0.8951 (>0.05) |
| Gamma (30 − 80 Hz) | p=0.1410 (>0.05) | p=0.7088 (>0.05) |

One important point to note that all the bivariate synchronization measures considered in this work, except h-square are symmetric, that is, synchronization between channels $i$ and $j$ is equal to synchronization between channels $j$ and $i$. So, we can form the symmetric $n \times n$ matrix ($n$ is the number of channels, across which synchronization is being measured) whose $ij$ th entry is the synchronization between channels $i$ and $j$. Since the matrix is real symmetric matrix, all its eigenvalues will be real. The highest eigenvalue of this matrix will give the dominant trend synchronization across all the $n$ channels (Schindler et al. 2007). Since h-square is not a symmetric measure, this method of extension to more than two channels will not work, and therefore we took average of the h-square values across all channel pairs in one direction (from channel $i$ to the channel $j$ for the synchronization between $i$ th channel and $j$ th channel). Obviously, it is a completely different extension method and therefore results of Table 4 (MI),

Table 5 (PS), Table 6 (AC) and Table 8 (coherence) are not comparable with those of Table 7 (h-square).

## 5. Discussion

Unlike the previous study (Arthuis et al. 2009), where patient's consciousness score was measured during seizure, we measured the same consciousness score immediately after the seizure offset (Table 2). This score was originally proposed in (Arthuis et al. 2009) and a score of 3 or more means impaired consciousness. The higher the score is greater is the impairment or loss of consciousness (LOC). The very first seizure is simple partial and therefore the score is zero (Table 2). Out of remaining 22 complex partial seizures the score at the post ictal period is either 1 or 2 only for the 4 out of 22 ($\approx 18\%$) and this indicates there was no measurable LOC for these 4 seizures after the offset. For remaining 82% of the seizures there was LOC after the seizure offset.

Our study of LOC associated with complex partial seizure is different in several ways than the precious study (Arthuis et al. 2009). First major difference is methodological. (A) They measured synchronization by h-square, which is a nonlinear, asymmetric correlation measure, whereas we measured synchronization by mutual information (MI), which is a symmetric, bivariate measure. (B) Their synchronization measure was bivariate and was carried out across only a few channel pairs, whose choice may be biased, whereas we extended the bivariate MI to multivariate MI and applied it on all the seizure onset zone (SOZ) channels and also on least involved in seizure channels. Total number of channels in our data from 11 patients was between 150 and 240 and therefore it was not possible to compute all the pairs. (C) We compared MI with all major synchronization measures in neuroscience, which showed that MI gives most statistically significant measure among all of them except h-square, the measure used in (Arthuis et al. 2009). However, multidimensional extension of h-square and MI are totally different and their statistical significance values in Table 6 and Table 3 respectively are not comparable to each other. (D) Unlike in (Arthuis et al. 2009) we have measured the synchronization in different frequency bands that have significance in cognition namely, delta, theta, alpha, beta and gamma. Statistically significant synchronization was observed across the SOZ channels only in beta and gamma range by MI and in gamma range also by phase synchronization (PS). h-square gives higher significant synchronization at beta and gamma band and even shows synchronization at alpha band. No other measure showed statistically significant synchronization across all the seizures at alpha band. h-square tends to give high value of synchronization compared to other measures (see Fig. 1 for a bivariate implementation), which might affect the results reported in (Arthuis et al. 2009). (E) (Arthuis et al. 2009) assessed consciousness during complex partial seizures, whereas we assessed it immediately after the offset, but reached at similar conclusions. This can be considered an extension of the earlier work. (F) Our study remained confined within temporal lobe. Unlike (Arthuis et al. 2009) we could not include thalamus and parietal cortex for clinical constraints. But since synchronization across the SOZ channels was quite significant in beta and gamma range, they likely have spread to wider brain areas (Kopell et al. 2000), supporting the study of (Arthuis et al. 2009). Since beta and gamma band EEG are more important for wide ranging cognitive tasks including sensory awareness, higher than usual synchronization in them is likely to lead to LOC in different degrees, which is evident from Table 2.

If synchronization is due to AC or PS or coherence or due to any other measure, MI is likely to reflect that synchronization, but not the other way round. In other words, if PS or AC is high MI will also have an upward trend, but if MI is high PS or AC may not show any increasing trend at all, because MI may be high due to some other synchronization measure(s). This is the reason the numerical value to MI is lower than other measures (Fig. 1). MI maintains a more stable trend (that is, less jittery) compared to other synchronization measures (Fig. 1). In short, the measure of synchronization given by MI is more generalized and robust than many other widely used synchronization measure used in neuroscience. If this is coupled with the fact that semantic information in a discrete signal is embedded in terms of 3-point motifs (Majumdar and Jayachandran 2018) then the way we have calculated the MI by frequency distribution of combination of motifs across two signals is a better measure of interdependence of the two signals than possible by PS or AC or coherence or some other measure alone.

It has been reported that, slow delta wave activity (1 – 2 Hz) in the bilateral frontal and parietal cortices during complex-partial seizures coupled with unilateral mesial temporal fast seizure activity spreading to the bilateral temporal lobes, might be responsible for impaired consciousness (Englot et al. 2010). Higher synchronization across the channels in temporal lobe in beta and gamma range during the latter half of complex partial seizure is indicative of the propagation of the fast seizure activity in the temporal lobe. However, by all synchronization measures that we studied in this work the delta band synchronization in temporal lobe remained insignificant (Tables 4 through 8). This means there is no delta wave propagation from temporal lobe to outside. The delta wave observed in temporal and parietal cortices are not related to the temporal lobe seizures.

One important finding is mutual information (which is a form of synchronization, see Fig. 1) across the seizure onset zone channels is more towards the seizure offset than any other duration, whereas mutual information is more either at the beginning or at the middle of seizure duration across the least involved channels (Fig. 5). This trend has been observed in the gamma band in 17 out of 22, that is, in 77% of the complex partial seizures that we studied (see the Supplement). The same trend, to a lesser extent, has been observed in beta band too (Fig. 5 and Supplement). In most of the cases mutual information across the least involved channels increased initially and as the seizure evolved towards the termination and post ictal state there was a steady increase in mutual information over the seizure onset zone channels (Fig. 5 and Supplement). High synchronization leading to seizure termination has already been observed in multiple studies (Schindler et al. 2007, Jiruska et al. 2013, Prasad et al. 2013, Majumdar et al. 2014). Here we have shown that high synchronization persists often beyond termination from least involved channels to seizure onset zone in beta and gamma bands and impairs consciousness.

We have used here to assess the state of consciousness of the patients according to the consciousness scoring scheme introduced by (Arthuis et al. 2009). However, there seems to be no straightforward (or linear, to be more precise) relationship between the level of synchronization and the level of consciousness. It was reported that excessive synchronization across wide brain regions during complex partial seizure is responsible for loss of consciousness (Arthuis et al. 2009). But if synchronization is high the loss of consciousness may not be as high and also the other way round as can be verified from the Supplement, where we have given all the mutual information measure across the seizure onset zone channels as well as the least involved channels before, during and after all the 23 seizures along with the consciousness score of the patient after the seizure offset. For example, seizures 21 and 22 are from the same patient with the same electrode implantation. The amount of mutual information across the seizure onset zone channels and also across the least involved channels in beta and gamma band is always less than 4 for seizure 21 (Supplement) with a consciousness score 6 after seizure offset, but for seizure 22 the mutual information in the same channels and in the same bands for the same duration is almost always higher than 4 and can be as high as 20, but with a consciousness score 4 (Supplement). This means loss of consciousness is more after seizure 21 than after seizure 22.

Among all the synchronization measuring algorithms we studied in this work h-square is the most computer time intensive algorithm even in bivariate case. The other synchronization measures that we studied in this work are all symmetric in the bivariate case (and much less time intensive to compute than h-square) and therefore their pairwise values will fill a square matrix whose highest eigenvalue (eigenvalues are always real as the matrix is symmetric) will give the principal or dominating trend of the synchronization measure across all the channels. Since in this extension matrix inversion is involved the time complexity is quite significant. On the other hand, since h-square is not a symmetric measure, we cannot use the principal component method for extension to more than 2 channels. We simply took average of pairwise h-square values for all the channel pairs. Even then multidimensional extension of h-square is much more time consuming than the same for the other synchronization measures studied here.

**Conclusion**

Here we have proposed an algorithm for measuring mutual information between two discrete signals and then extended it to more than two signals. We have compared this measure with a number of other widely used bivariate signal dependency measures in neuroscience both for a pair of signals and for more than two signals and determined their statistical significance. We have demonstrated that mutual information is a robust, generalized signal dependency measure. Then we applied this and other synchronization measures namely, (Hilbert) phase synchronization, amplitude correlation, nonlinear correlation (h-square) and coherence on multichannel iEEG data consisting of 23 temporal lobe onset seizures recorded from 11 patients. Statistically significant synchronization across the seizure onset zone was found in beta and gamma range during the latter part of complex partial seizures in conjunction with impaired consciousness (determined for 77% of the complex partial seizures) after the offset in 82% of them (Supplement). Our data shows that there is reasonable association between higher level of synchronization and loss of consciousness as reported in (Arthuis et al. 2009).

However, the relationship between the level of synchronization and the amount of loss of consciousness is far from obvious. More elaborate investigation needs to be undertaken in this regard recruiting brain areas like thalamus and parietal cortex, which are implicated more in consciousness than temporal lobe. The generalized mutual information that we developed here can be applied on more than two channels sampled from cortical areas implicated in consciousness to determine the overall level of mutual information across those regions during and after complex partial seizures. Then attempt can be made to relate the ensemble mutual information at different frequency bands to the level of loss of consciousness according to consciousness score to extend our current work and the work reported in (Arthuis et al. 2009). It will also worth investigating the relationship between delta band ensemble mutual information in frontal and parietal cortices and beta and gamma band mutual information in unilateral temporal lobe seizure onset zone with respect to the loss of consciousness. This will be an extension of the work reported in (Englot et al. 2010).

Any discrete signal can be decomposed into 13 different 3-point motifs, each of which is encoding information into that signal in the form of its shape (Majumdar and Jayachandran 2018). Mutual information between such representations of signals is likely to be more meaningful and informative.

**Acknowledgement**

This work was supported by the Department of Biotechnology, Government of India grant no. BT/PR7666/MED/30/936/2013 and National Science Foundation (USA) award no. NSF RII-2 FEC OIA 1632891.

**References**

M. Arthuis, L. Valton, J. Regis, P. Chauvel, F. Wendling, L. Naccache, C. Bernard and F. Bartolomei, 2009. Impaired consciousness during temporal lobe seizures is related to increased long-distance cortico-subcortical synchronization, Brain 132, 2091–2101.

S. Aydore, D. Pantazis and R. M. Lehay, 2013. A note on phase locking value and its properties, NeuroImage 74, 231–244.

C. Babiloni, F. Vecchio, G. Mirabella, M. Buttiglione, F. Sabastiano, A. Picardi, G. Di Gennaro, P. P. Quarato, L. G. Grammaldo, P. Buffo, V. Esposito, M. Manfredi, G. Cantore and F. Eusebi, 2009. Hippocampal, amygdala and neocortical synchronization of theta rhythms is related to an immediate recall during Rey auditory verbal learning test, Hum. Brain Map. 30, 2077–2089.

E. Basar, C. Basar-Eroglu, S. Karakas and M. Schurmann, 2001. Gamma, alpha, delta and theta oscillations govern cognitive processes, Int. J. Psychophysiol. 39, 241–248.

A. M. Bastos, J.-M. Schoffelen, 2016. A tutorial review of functional connectivity analysis methods and their interpretational pitfalls, Front. Syst. Neurosci., 9, 175. doi: 10.3389/fnsys.2015.00175.

H. Blumenfeld and J. Taylor, 2003. Why do seizures cause loss of consciousness? Neuroscientist 9, 301–310.

T. M. Cover and J. A. Thomas, 2006. Elements of Information Theory, 2nd ed., John Wiley & Sons, Hoboken, NJ, USA.

F. Crick and C. Koch, 1990. Some reflections on visual awareness, Cold Spring Harbor Symp. Quant. Biol. 55, 953–962.

J. Engel Jr., P. C. Van Ness, T. B. Rasmussen, L. M. Ojemann, 1993. Outcome with respect to epileptic seizures, In: J. Engel Jr. (ed.) Surgical Treatment of the Epilepsies, 2nd ed. Raven Press, New York, 609–621.

D. J. Englot, L. Yang, H. Hamid, N. Danielson, X. Bai, A. Marfeo, L. Yu, A. Gordon, M. J. Purcaro, J. E. Motelow, R. Agarwal, D. J. Ellens, J. D. Golomb, M. C. F. Shamy, H. Zhang, C. Carlson, W. Doyle, O. Devinsky, K. Vives, D. D. Spencer, S. S. Spencer, C. Schevon, H. P. Zaveri and H. Blumenfeld, 2010. Impaired consciousness in temporal lobe seizures: role of cortical slow activity, Brain 133, 3764–3777.

J. Fell and N. Axmacher, 2011. The role of phase synchronization in memory processes, Nat. Rev. Neurosci. 12, 105–118.

P. Fries, 2009. Neuronal gamma-band synchronization as a fundamental process of cortical computation, Ann. Rev. Neurosci. 32, 209–224.

B. Guntekin and E. Basar, 2016. Review of evoked and event-related delta responses in the human brain, Int. J. Psychophysiol. 103, 43–52.

P. Jiruska, M. de Curtis, J. G. R. Jefferys, C. A. Schevon, S. J. Schiff and K. Schindler, 2013. Synchronization and desynchronization in epilepsy: controversies and hypothesises, J. Physiol. 591, 787–797.

G. G. Knyazev, J. Y. Slobodskoj-Plusnin and A. V. Bocharov, 2009. Event-related delta and theta synchronization during explicit and implicit emotion processing, Neurosci. 164, 1588–1600.

N. Kopell, G. B. Ermentraut, M. A. Whittington, R. D. Traub, 2000. Gamma rhythms and beta rhythms have different synchronization properties, PNAS. 97: 1867–1872.

X. Li and G. Ouyang, 2010. Estimating coupling direction between neuronal populations with permutation conditional mutual information, NeuroImage 52, 497–507.

J. Lin, E. Keogh, S. Lonardi and P. Patel, 2002. Finding motifs in time series, Proc. SIGKDD'02, the 8th ACM workshop on knowledge discovery and data mining, Edmonton, Alberta, Canada, 53–68.

K. Majumdar, P. D. Prasad and S. Verma, 2014. Synchronization implies seizure or seizure implies synchronization? Brain. Topogr. 27, 112–122.

K. Majumdar and S. Jayachandran, 2018. A geometric analysis of time series leading to information encoding and a new entropy measure, J. Comp. Appl. Math. 328, 469–484.

A. R. McIntosh, M. N. Rajah and N. J. Lobaugh, 1999. Interactions of prefrontal cortex in relation to awareness in sensory learning, Science 284, 1531–1533.

L. Melloni, C. Molina, M. Pena, D. Torres, W. Singer and E. Rodriguez, 2007. Synchronization of neural activity across cortical areas correlates with conscious perception, J. Neurosci. 27, 2858–2865.

F. Mormann, K. Lehnertz, P. David and C. E. Elger, 2000. Mean phase coherence as a measure for phase synchronization and its application to the EEG of epilepsy patients, Physica D 144, 358–369.

E. Olofsen, J. W. Sleigh and A. Dahan, 2008. Permutation entropy of the electroencephalogram: a measure of anaesthetic drug effect, Brit. J. Anasthes. 101, 810–821.

P. D. Prasad, S. V. Datta, K. Majumdar, 2013. Enhanced phase and amplitude synchronization towards focal seizure offset, Clin. EEG. Neurosci. 44, 16–24.

E. Rodriguez, N. George, J.-P. Lachaux, J. Martinerie, B. Renault and F. J. Varela, 1999. Perception's shadow: long-distance synchronization of human brain activity, Nature 397, 430–433.

K. Schindler, H. Leung, C. E. Elger and K. Lehnertz, 2007. Assessing seizure dynamics by analysing the correlation structure of multichannel intracranial EEG, Brain 130, 65–77.

B. Spitzer and S. Haegens, 2017. Beyond the status quo : a role for beta oscillations in endogenous content (re)activation, eNeuro 4(4), 1–15.

P. N. Steinmetz, A. Roy, P. J. Fitzgerald, S. S. Hsiao, K. O. Johnson and E. Neibur, 2000. Attention modulates synchronized neuronal firing in primate somatosensory cortex, Nature 404, 187–190.

C. Tallon-Baudry, 2009. The roles of gamma-band oscillatory synchrony in human visual cognition, Front. Biosci. 14, 321–332.

L. M. Ward, 2003. Synchronous neuronal oscillations and cognitive processes, Trends Cog. Sci. 7, 553–559.

F. Wendling, F. Bartolomei, J. J. Bellanger and P. Cauvel, 2001. Interpretation of interdependencies in epileptic signals using a macroscopic physiological model of the EEG, Clin. Neurophysiol. 112, 1201–1218.